\documentclass[conference]{IEEEtran}
\IEEEoverridecommandlockouts
\usepackage{cite}
\usepackage{amsmath,amssymb,amsfonts}
\usepackage{algorithmic}
\usepackage{graphicx}
\usepackage{textcomp}
\usepackage{xcolor}
\def\BibTeX{{\rm B\kern-.05em{\sc i\kern-.025em b}\kern-.08em
    T\kern-.1667em\lower.7ex\hbox{E}\kern-.125emX}}

\usepackage{microtype}
\usepackage{graphicx}
\usepackage{subfigure}
\usepackage{booktabs} % for professional tables

\newtheorem{theorem}{Theorem}[section]
\newtheorem{proposition}[theorem]{Proposition}
\newtheorem{lemma}[theorem]{Lemma}
\newtheorem{corollary}[theorem]{Corollary}
\newtheorem{definition}[theorem]{Definition}

\usepackage{algorithm}
\usepackage{algorithmic}

\begin{document}

\title{
LiSFC-Search: Lifelong Search for Network SFC Optimization under Non-stationary Drifts

}

\author{\IEEEauthorblockN{1\textsuperscript{st} Zuyuan Zhang}
\IEEEauthorblockA{\textit{The George Washington University} \\
zuyuan.zhang@gwu.edu}
\and
\IEEEauthorblockN{2\textsuperscript{nd} Vaneet Aggarwal}
\IEEEauthorblockA{\textit{Purdue University} \\
vaneet@purdue.edu}
\and
\IEEEauthorblockN{3\textsuperscript{rd} Tian Lan}
\IEEEauthorblockA{\textit{The George Washington University} \\
tlan@gwu.edu}
}

\maketitle

\begin{abstract}
Edge–cloud convergence is reshaping service provisioning across 5G/6G and computing power networks (CPNs). Service function chaining (SFC) requires continuously placing and scheduling virtual network functions (VNFs) chains under compute/bandwidth and end-to-end QoS constraints. Most SFC optimizers assume static or stationary networks, and degrade under long-term topology/resource changes (failures, upgrades, expansions) that induce non-stationary graph drifts. We propose LiSFC, a Lipschitz lifelong planner that transfers MCTS statistics across drifting network configurations using an MDP-distance bound. More precisely, we formulate the problem as a sequence of MDPs indexed by the underlying network graph and constraints, and we define a \emph{graph drift} metric that upper-bounds the LiZero MDP distance. This allows LiSFC to import theoretical guarantees on bias and sample efficiency from the LiZero framework while being tailored to cloud–network convergence. We then design \emph{LiSFC-Search}, an SFC-aware unified MCTS (UMCTS) procedure that uses transferable adaptive UCT (aUCT) bonuses to reuse search statistics from prior CPN configurations. Preliminary results on synthetic CPN topologies and SFC workloads show that LiSFC consistently reduces SFC blocking probability and improves tail delay compared to non-transfer MCTS and purely learning-based baselines, highlighting its potential as an AI/ML building block for cloud–network convergence.
\end{abstract}

\section{Introduction}
The Cloud–Network Convergence (CNC) vision aims to tightly integrate cloud computing, computing power networks (CPNs), and heterogeneous wireless/optical infrastructures in order to deliver low-latency, high-bandwidth, and highly reliable networked services at scale~\cite{yukun2024computing,liu2022computing,ma2023research,tang2021computing,yan2024becs,zou2024distributed,qiao2024br,zhang2024distributed}. In this paradigm, network resources (links, VNFs, servers) and cloud resources (CPUs, GPUs, storage) are orchestrated as a unified substrate, enabling on-demand deployment of network services as software-defined, cloud-native microservices~\cite{adoga2022network,bhamare2016survey,xie1608service}. \emph{Service function chaining} (SFC) is a key abstraction in this setting: traffic flows are steered through ordered sequences of virtual network functions (VNFs)—such as firewalls, encoders, and application-layer middleboxes—that may be instantiated across multiple edge and cloud sites~\cite{bhamare2016survey,kuo2018deploying,jin2020latency}. The ability to jointly optimize SFC placement and scheduling (which is defined as our SFC optimization in this paper) is crucial for supporting this vision.

Despite significant progress, SFC placing–scheduling 
%in converged cloud–network environments 
remains challenging for three reasons~\cite{adoga2022network,kuo2018deploying,sallam2018shortest,jin2020latency,shang2020online,shang2020online}. \emph{First}, the underlying cloud–network graph is inherently \emph{non-stationary}: links and servers may be added, upgraded, or fail; traffic patterns evolve; and new service classes are introduced over time. This graph drift breaks the assumptions of classical one-shot optimization models, such as mixed integer linear programs (MILPs), which are typically solved per snapshot and cannot amortize computation across network evolution. \emph{Second}, SFC planning is multi-dimensional: each decision must jointly consider compute capacity, link bandwidth, and end-to-end (E2E) latency constraints across multiple timescales (short-term flow scheduling vs.\ long-term VNF placement), which makes purely reactive heuristics brittle under load spikes and topology changes. \emph{Third}, while AI/ML has been increasingly applied to SFC and NFV (e.g., deep reinforcement learning or imitation-based policies), most existing methods are either trained for a single topology or retrained from scratch when the network changes, leading to slow adaptation and high sample complexity in CNC-scale systems~\cite{xiao2019nfvdeep,liu2024service,polverini2020improving,qu2021reliable,kong2024deep,feng2023drl}.

This paper proposes \emph{LiSFC}, a \emph{Lipschitz Lifelong SFC planning} framework . We first cast dynamic SFC placing–scheduling in a CPN as a \emph{lifelong task family} of MDPs, where each task corresponds to a particular cloud–network graph and workload regime. Building on LiZero formulation, we introduce an SFC-specific \emph{graph drift} metric that aggregates spectral, capacity, and bandwidth changes of the underlying CPN. We then show that this graph drift upper-bounds the LiZero MDP distance for the SFC MDP, which means that the Lipschitz and bias bounds established in~\cite{zhang2025lipschitz} can be then applied (relying on our notion of graph drift) in the LiSFC setting. Intuitively, small changes in topology or capacity lead to proportionally small perturbations in transition dynamics and rewards, so previously learned Q-values and visit counts remain informative under moderate graph drift. The proposed LiSFC enables efficient lifelong search for optimal SFC placement and scheduling under non-stationary drifts.

On top of this modeling layer, we design \emph{LiSFC-Search}, a unified MCTS (UMCTS) procedure specialized for SFC planning in converged cloud–network infrastructures. LiSFC-Search combines two ingredients: (i) it uses \emph{transferable aUCT bonuses} derived from LiZero to safely reuse MCTS statistics collected on previous CPN configurations, with exploration penalties scaled by the estimated graph drift; and (ii) it employs an importance-weighted estimator of the SFC MDP distance that can be computed from offline logs of SFC requests and network states. Conceptually, LiZero supplies the lifelong planning backbone that makes this prior robust under non-stationary CNC environments.

Our preliminary evaluation on synthetic CPN topologies and SFC workloads shows that LiSFC enjoys the best of both worlds. Compared to running Greedy algorithm alone, LiSFC-Search improves robustness under topology and load shifts by actively exploring alternative placements when the demonstration prior becomes stale. Compared to a standard, non-transfer MCTS planner, LiSFC drastically reduces search depth and simulation count needed to reach high-quality decisions, especially when the new CPN configuration is close (in graph drift) to previously seen ones. As a result, LiSFC achieves lower SFC blocking probability and improved tail E2E delay across a range of CNC-relevant scenarios. These results suggest that Lipschitz lifelong planning, when combined with demonstration learning, is a promising direction for AI/ML-driven orchestration in cloud–network convergence.

\paragraph*{Contributions.}
In summary, this paper makes the following contributions:
\begin{itemize}
  \item We formulate dynamic SFC placing–scheduling as a \emph{lifelong} learning problem over a sequence of varying MDPs, explicitly modeling cloud–network graph drift and connecting it to LiZero's MDP distance.
  \item We introduce LiSFC, a Lipschitz lifelong SFC planner that uses LiZero's transferable MCTS and aUCT rule~\cite{zhang2025lipschitz}.
  
  \item We design LiSFC-Search, an SFC-aware unified MCTS procedure that injects Network Diffuser subtrees as demonstrations and reuses MCTS statistics across CPN configurations using a graph-drift-based distance estimator.
  \item We provide preliminary empirical evidence that LiSFC improves both SFC blocking and E2E delay under non-stationary CNC scenarios compared to non-transfer MCTS and learning-only baselines, highlighting its potential as an AI/ML primitive for cloud–network convergence.
\end{itemize}

\section{Preliminaries and Problem Setup}

This section connects the SFC optimization model with the Lipschitz lifelong MCTS framework in~\cite{zhang2025lipschitz}. We first recall the network and SFC model, then cast the online placing--scheduling problem as an MDP, and finally define a family of related SFC tasks that share structure but differ in the underlying network graph. This will let us reuse the LiZero theory while tailoring it to the SFC setting.

\subsection{Network and SFC Model}

\begin{definition}[Network and SFC]
Let the physical network be an undirected graph \(G=(V,E)\) with node capacities \(\{C^V_j\}_{j\in V}\) and link bandwidths \(\{B^E_{pq}\}_{(p,q)\in E}\). Each SFC request \(i\) is a chain of VNFs indexed by \(j\), where the \(j\)-th VNF requires compute demand \(c_{i,j}\) on some node, and the corresponding flow between VNF \(j\) and \(j{+}1\) requires bandwidth demand \(b_{i,j}\) on some path. Each SFC has a release time, processing duration, and deadline. A joint placing--scheduling decision is feasible if, at every time slot, the total VNF demand on each node does not exceed its capacity and the total flow demand on each link does not exceed its bandwidth.
\end{definition}

\subsection{SFC Planning MDP}

\begin{definition}[SFC MDP]
We model online SFC optimization as an MDP \(\mathcal{M}=\langle\mathcal{S},\mathcal{A},P,r,\gamma\rangle\). The state \(s\in\mathcal{S}\) encodes the current network resources and the queue of active/waiting SFCs (residual capacities, remaining durations, deadlines, etc.). The action \(a\in\mathcal{A}\) decides SFC admission, placement and activation subject to feasibility. The transition kernel \(P(\cdot\mid s,a)\) follows capacity accounting and SFC arrivals/completions. The reward \(r(s,a)\) balances SFC completion (success), delay, and blocking penalties, and \(\gamma\in(0,1)\) is the discount factor.
\end{definition}

Thus, each SFC environment is an MDP with the same state and action spaces but possibly different transition and reward functions, depending on the graph and traffic statistics.

\subsection{Lifelong Task Family}

\begin{definition}[Lifelong sequence]
A lifelong SFC setting is a sequence of related MDPs
\(
\{\mathcal{M}_1,\dots,\mathcal{M}_m,\mathcal{M}\}
\),
where each \(\mathcal{M}_i\) corresponds to an SFC environment on a historical network snapshot, and \(\mathcal{M}\) is the new task defined on a perturbed graph \(G'\). The tasks share \((\mathcal{S},\mathcal{A},\gamma)\) but differ in \((P_i,r_i)\) through changes in topology, capacities, and bandwidths.
\end{definition}

Our goal is to reuse search knowledge from \(\{\mathcal{M}_i\}\) when planning in \(\mathcal{M}\), while controlling the error in terms of how much the underlying networks have changed.

\section{Graph Drift and MDP Distance}
\label{sec:graph-drift}
This section provides the bridge between network-level changes and decision-level changes. We first define a graph drift metric \(\Delta G\) that aggregates spectral, capacity, bandwidth and (optionally) edit differences between two networks. We then recall the Lipschitz MDP distance \(d(M,M')\) from~\cite{zhang2025lipschitz} and show that, under mild Lipschitz assumptions tailored to SFC planning, \(\Delta G\) upper-bounds \(d(M,M')\). Combining this with the aUCT analysis of~\cite{zhang2025lipschitz} yields decision-level transfer bounds that depend only on network statistics and graph drift, rather than on an abstract MDP distance.

\begin{definition}[Graph drift \(\Delta G\)]
Let \(G=(V,E)\) and \(G'=(V',E')\) be two network graphs with capacities and bandwidths \(\{C^V_j,B^E_{pq}\}\) and \(\{{C'}^V_j,{B'}^E_{pq}\}\). Let \(\Delta_{\text{spec}},\Delta_{\text{cap}},\Delta_{\text{bw}},\Delta_{\text{edit}}\) denote, respectively, a spectral distance between \(G\) and \(G'\), the aggregate change in node capacities, the aggregate change in link bandwidths, and an (optional) graph edit distance. For nonnegative weights \(w_k\), define
\[
\Delta G = \sum_k w_k \Delta_k,
\quad
\Delta_k\in\{\Delta_{\text{spec}},\Delta_{\text{cap}},\Delta_{\text{bw}},\Delta_{\text{edit}}\}.
\]
\end{definition}

The exact form of each component can follow the choices in~\cite{zhang2025network}; the key point is that \(\Delta G\) summarizes how much routing options and resource margins differ between two networks, and is directly computable from network-level statistics.

\begin{definition}[MDP distance \(d(M,M')\)]
Following~\cite{zhang2025lipschitz}, for two MDPs \(M=\langle R,P\rangle\) and \(M'=\langle R',P'\rangle\) sharing state/action spaces and a common sampling measure \(U\) on triples \((s,a,s')\), we define
\[
d(M,M')
=
\mathbb{E}_{(s,a,s')\sim U}
\big[
\lvert R^a_s-R^{\prime a}_s\rvert + \kappa\,\lvert P^a_{ss'}-P^{\prime a}_{ss'}\rvert
\big],
\]
where \(\kappa>0\) trades off reward and transition differences.
\end{definition}

Intuitively, \(d(M,M')\) is the expected one-step discrepancy between the two MDPs under a reference distribution \(U\) (e.g., uniform or induced by a mixture of policies).

\begin{lemma}[Graph drift as a proxy for MDP distance]
\label{lem:DG2d}
Assume that rewards and transitions in the SFC MDPs are Lipschitz with respect to resource margins and connectivity: small changes in capacities/bandwidths and topology induce changes in \(R\) and \(P\) bounded linearly by \(\Delta_{\text{cap}},\Delta_{\text{bw}},\Delta_{\text{spec}},\Delta_{\text{edit}}\). Then there exists \(c>0\) such that
\[
d(M,M') \le c\,\Delta G.
\]
\end{lemma}

\noindent
The proof follows the same Lipschitz argument as in~\cite{zhang2025network,zhang2025lipschitz} and is omitted due to space. Conceptually, \(\Delta G\) acts as a computable surrogate that upper-bounds the MDP distance. This allows us to plug \(\Delta G\) directly into the aUCT bounds of~\cite{zhang2025lipschitz} and control transfer error using only network statistics.

\section{Transferable aUCT for Lifelong Search}
We now recall how LiZero~\cite{zhang2025lipschitz} uses the MDP distance \(d(M,M')\) to build a Lipschitz upper confidence rule for MCTS, and then specialize it to the SFC setting using graph drift. We first state the Lipschitz aUCT bound that relates the empirical Q-values of two MDPs to their distance and visit counts. Combining this with Lemma~\ref{lem:DG2d} yields a graph-drift version in which the transfer bias term depends only on \(\Delta G\). Finally, we derive a transferable UCB over past SFC tasks. All proofs are identical to~\cite{zhang2025lipschitz} and omitted.

\begin{theorem}[Lipschitz aUCT~{\cite[Thm.~3.2]{zhang2025lipschitz}}]
\label{thm:auct}
Let \(M,M'\) be two MDPs sharing state and action spaces, with discount \(\gamma\in(0,1)\) and rewards bounded by \(|r|\le R_{\max}\). Let \(Q_M^N(s,a)\) and \(Q_{M'}^{N'}(s,a)\) be the empirical Q-values estimated by UCT at edge \((s,a)\) after \(N\) and \(N'\) visits, respectively. Then, for any \(\delta\in(0,1)\), with probability at least \(1-\delta\),
\begin{align*}
\big\lvert Q_M^N(s,a)-&Q_{M'}^{N'}(s,a)\big\rvert
\le \\
&\frac{1}{1-\gamma}\,d(M,M')
+
\frac{2R_{\max}}{1-\gamma}
\sqrt{\frac{\ln(2/\delta)}{2\min\{N,N'\}}}.    
\end{align*}
\end{theorem}

The first term depends only on the MDP distance \(d(M,M')\) and controls the transfer bias, while the second term is a standard Hoeffding-type concentration term that shrinks as visit counts grow.

\begin{corollary}[Graph-drift aUCT bound]
\label{cor:auct-deltaG}
Under the assumptions of Lemma~\ref{lem:DG2d}, the bound in Theorem~\ref{thm:auct} can be rewritten as
\begin{align*}
\big\lvert Q_M^N(s,a)-&Q_{M'}^{N'}(s,a)\big\rvert
\le\\
&\frac{c}{1-\gamma}\,\Delta G(M,M')
+
\frac{2R_{\max}}{1-\gamma}
\sqrt{\frac{\ln(2/\delta)}{2\min\{N,N'\}}}. 
\end{align*}
In particular, the transfer bias term is now controlled directly by the graph drift \(\Delta G(M,M')\), which depends only on network-level statistics.
\end{corollary}

We next adapt the multi-task transfer bound of LiZero to the SFC setting and express it entirely in terms of graph drift.

\begin{corollary}[Graph-drift Transfer UCB]
\label{cor:transfer-ucb-deltaG}
Let \(\mathcal{M}_1,\dots,\mathcal{M}_m\) be prior SFC MDPs and \(\mathcal{M}\) be the new task. For each prior task \(\mathcal{M}_i\), let \(Q_{\mathcal{M}_i}^{N_i}(s,a)\) be the empirical Q-value at \((s,a)\) after \(N_i(s,a)\) visits. Then, under the assumptions of Lemma~\ref{lem:DG2d}, for any \(\delta\in(0,1)\), with probability at least \(1-\delta\),

\begin{align*}
Q^*_{\mathcal{M}}(s,a)
\le \min_{i\le m}
\Big[
Q_{\mathcal{M}_i}^{N_i}(s,a)
+
& \frac{c}{1-\gamma}\,\Delta  G(\mathcal{M},\mathcal{M}_i)
\\& +
\frac{2R_{\max}}{1-\gamma}
\sqrt{\frac{\ln(2/\delta)}{2N_i(s,a)}}
\Big].  
% \vspace{-1cm}
\end{align*}

\end{corollary}

This is the transferable upper confidence term we will actually use inside LiSFC-Search: for each \((s,a)\), we take the minimum over all bias-corrected UCBs from prior tasks, where the bias is expressed solely in terms of graph drift.

\section{Algorithm: LiSFC-Search}
Building on the above theory, LiSFC-Search combines standard UCT with graph-drift-based transferable aUCT bounds and network-aware subtree reuse. The planner maintains a knowledge base of subtrees and estimated graph drifts between tasks, injects relevant subtrees into the current search tree, and uses a max--min selection rule that balances exploitation of the current task and safe transfer from prior tasks.

\begin{algorithm}[t]
\caption{LiSFC-Search (UMCTS with graph-drift transferable aUCT)}
\label{alg:lisfc}
\begin{algorithmic}[1]
\STATE \textbf{Inputs:} prior tasks \(\{\mathcal{M}_i\}\), new task \(\mathcal{M}\) with graph \(G\), knowledge base \(\mathcal{K}\).
\FOR{each episode}
  \STATE Set root state \(s\gets\) current SFC MDP state.
  \STATE \texttt{InjectSubtrees}(\(\mathcal{K},G\)) to attach subtrees from prior tasks with small graph drift \(\Delta G(\mathcal{M},\mathcal{M}_i)\).
  \FOR{simulation index \(\mathrm{sim}=1,\dots,S\)}
    \STATE From the root, descend by selecting
    \begin{align*}
        a = \arg\max_a
      \min\Big\{
        \hat Q(s,a) + C&\sqrt{\tfrac{\ln N(s)}{N(s,a)}},\\ & U_{\text{aUCT}}(s,a)
      \Big\},
    \end{align*}
    where \(\hat Q(s,a)\) is the empirical value, \(N(s)\) and \(N(s,a)\) are visit counts, and \(U_{\text{aUCT}}(s,a)\) is the graph-drift-based upper confidence bound given by Corollary~\ref{cor:transfer-ucb-deltaG}.
    \STATE Expand a feasible child by applying \(a\) (respecting SFC resource constraints), perform a rollout with a default policy, and backpropagate the return.
  \ENDFOR
  \STATE Execute the best admissible root action in the real SFC system and update \(\mathcal{K}\) with the explored subtree and updated graph drift estimates.
\ENDFOR
\end{algorithmic}
\end{algorithm}

When the new SFC task is very similar to past tasks (small \(\Delta G\) and hence small MDP distance), LiSFC-Search aggressively reuses prior subtrees and relies on tight transferable UCBs; when all tasks are far, the algorithm naturally falls back to standard UCT behavior.

\section{Estimating Distances}
To make the aUCT bounds practical, we need quantities that upper-bound \(d(M,M')\) in a way that is both theoretically sound and computationally convenient. Lemma~\ref{lem:DG2d} already provides a purely graph-based surrogate via \(\Delta G\). In addition, LiZero~\cite{zhang2025lipschitz} proposes an importance sampling estimator for \(d(M,M')\) based on rollouts; LiSFC can use either the coarse, structure-aware upper bound \(c\,\Delta G\), or, when enough trajectories are available, plug the empirical estimate of \(d\) into the same graph-drift aUCT template.

For completeness, we restate the IS estimator specialized to SFC transitions and refer to~\cite[Section~4]{zhang2025lipschitz} for proofs.

\begin{definition}[IS estimator of \(d\)]
Let \((s_i,a_i,s'_i)\) be samples collected under a behavior distribution \(\pi(s,a,s')\) during search. Define one-step discrepancies \(\Delta R_i = \lvert R^a_s - R^{\prime a}_s\rvert\) and \(\Delta P_i = \lvert P^a_{ss'} - P^{\prime a}_{ss'}\rvert\) evaluated at \((s_i,a_i,s'_i)\). For the reference measure \(U\) in Definition~\ref{lem:DG2d}, the importance sampling estimator is
\[
\widehat d
=
\frac{1}{n}\sum_{i=1}^n
w_i\big(\Delta R_i + \kappa\Delta P_i\big),
\quad
w_i = \frac{U(s_i,a_i,s'_i)}{\pi(s_i,a_i,s'_i)}.
\]
\end{definition}

Under standard coverage and boundedness assumptions, \(\widehat d\) is unbiased and converges to \(d(M,M')\) with a finite-sample concentration guarantee; the exact sampling complexity is given by Theorem~4.1 in~\cite{zhang2025lipschitz}. In LiSFC, we treat \(c\,\Delta G\) as a safe prior upper bound on \(d\), and optionally refine it with \(\widehat d\) when sufficient data is available.

\section{Complexity and Sample Efficiency}
Finally, we briefly discuss how graph-drift-aware transfer reduces the exploration cost of MCTS in SFC planning. Classical UCT must visit each action many times before elimination, even if it is clearly suboptimal in all related tasks. With the graph-drift Transfer UCB, LiSFC-Search can discard suboptimal actions much earlier when prior tasks with small \(\Delta G(\mathcal{M},\mathcal{M}_i)\) already provide tight upper bounds.

\begin{proposition}[Elimination under graph-drift transfer]
Consider a state \(s\) and a suboptimal action \(a\) in the new task \(\mathcal{M}\). Suppose there exists a prior task \(\mathcal{M}_i\) such that the transferable upper bound \(U_{\text{aUCT}}(s,a)\) from Corollary~\ref{cor:transfer-ucb-deltaG} falls below the lower confidence bound of the best action at \(s\) after \(N_i(s,a)\) visits in \(\mathcal{M}_i\). Then LiSFC-Search will eliminate \(a\) from further exploration at \(s\) after at most \(\tilde O(f(\Delta G(\mathcal{M},\mathcal{M}_i),\Delta_a))\) visits, where \(\Delta_a\) is the suboptimality gap and \(f(\cdot)\) decreases as the graph drift \(\Delta G(\mathcal{M},\mathcal{M}_i)\) decreases.
\end{proposition}

The precise form of \(f(\cdot)\) and the resulting sample complexity follow the same analysis as in~\cite[Sec.~3.3]{zhang2025lipschitz}. Here we emphasize the qualitative effect: closer SFC tasks (small graph drift) yield tighter transferable UCBs, earlier action elimination and hence fewer explorations, while distant tasks contribute little and the algorithm behaves like standard UCT.

\section{Experiments}
\label{sec:experiments}

In this section we provide a first empirical evaluation of LiSFC on dynamic CPN-style topologies and SFC workloads that reflect cloud--network convergence scenarios. Our goal is not to exhaustively tune all components, but to answer three concrete questions: (i) can LiSFC improve SFC-level performance metrics (blocking and end-to-end delay) compared to non-transfer MCTS and learning-only baselines under varying load; (ii) does LiSFC retain its advantages when the cloud--network graph drifts over time; and (iii) how does graph drift relate to the amount of search effort (simulations) LiSFC needs to adapt to new CPN configurations.

\subsection{Experimental Setup}
\label{sec:exp-setup}
\paragraph*{Topologies and traffic.}
We consider a medium-size CPN-style topology with \(|V|=20\) nodes and \(|E|=40\) links. Nodes are partitioned into access (edge) and core regions, with heterogeneous CPU capacities and link bandwidths that emulate converged cloud--network resources.
We generate a family of related graphs $\{G_0,G_1,G_2,G_3\}$: $G_0$ is the base topology, $G_1$ models a moderate upgrade, $G_2$ a degraded configuration, and $G_3$ a mixed perturbation. For each graph, we generate Poisson arrivals of SFC requests with 3--5 VNFs per chain, heterogeneous VNF types, and latency/deadline constraints. The offered load is varied by scaling the arrival rate.

\paragraph*{Algorithms.}
We compare the following methods:
\begin{itemize}
    \item \textbf{NF-Heuristic}: a conventional greedy SFC placer that selects shortest paths and packs VNFs based on residual capacity, similar to~\cite{zhang2025network}.
    \item \textbf{UMCTS (no transfer)}: a unified MCTS planner that uses standard UCT but does not reuse search statistics or graph drift; it plans independently on each \(G_k\).
    \item \textbf{LiSFC}: the proposed lifelong SFC planner, using LiSFC-Search (Algorithm~\ref{alg:lisfc}) with transferable aUCT and distance estimation and reusing statistics across \(\{G_0,G_1,G_2\}\).
\end{itemize}

\paragraph*{Metrics and implementation.}
We report (i) \emph{SFC blocking probability}, the fraction of SFC requests that are rejected due to capacity or deadline violations; (ii) \emph{95th percentile (P95) end-to-end delay} for accepted SFC flows; and (iii) \emph{average simulations per decision}, i.e., the average number of MCTS rollouts required by each planner to take one placement decision. All methods are implemented in the same discrete-event simulator, and for MCTS-based methods we use a fixed simulation budget per decision unless otherwise stated. Each experiment is repeated over multiple random seeds; error bars correspond to standard errors.

\subsection{Scenario~1: Load Variation on a Fixed CPN}

We first study how LiSFC behaves under increasing offered load on a fixed CPN topology \(G_0\). Here graph drift is zero, and the main question is whether LiSFC can perform well without any initial information.

\begin{figure}[t]
  \centering
  \includegraphics[width=0.95\columnwidth]{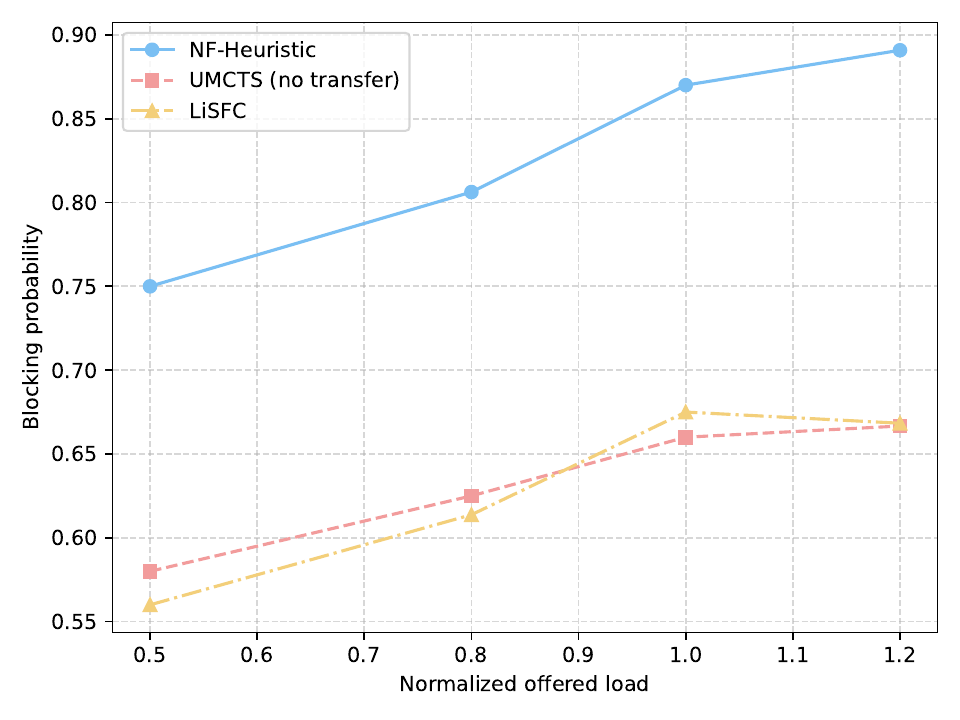}
  \caption{SFC blocking probability versus normalized offered load on the base CPN topology \(G_0\). }
  \label{fig:blocking}
  % \vspace{-0.5cm}
\end{figure}

Fig.~\ref{fig:blocking} shows the SFC blocking probability as a function of normalized offered load on \(G_0\). As expected, all methods exhibit increased blocking as the system approaches saturation. The greedy NF-Heuristic rejects many SFCs even at moderate load due to its myopic placement decisions. UMCTS (no transfer) and LiSFC can, in principle, approximate the optimal decision, and therefore, they consistently maintain a low blocking rate in the medium to high load range.

\subsection{Scenario~2: Graph Drift and Transfer Across CPN Configurations}

We next consider the full lifelong setting in which the cloud--network graph drifts over time from \(G_0\) to \(G_1\), \(G_2\) and \(G_3\). For each configuration, we reuse the same SFC workload model but adjust capacities and bandwidths according to the topology. Our main interest is how much LiSFC can reuse search statistics from \(G_0\) when planning in \(G_1\), \(G_2\) and \(G_3\), and how this relates to the graph drift \(\Delta G(G_0,G_k)\) and the MDP distance estimates \(\widehat d(\mathcal{M}_0,\mathcal{M}_k)\).

\begin{figure}[t]
  \centering
  \includegraphics[width=0.95\columnwidth]{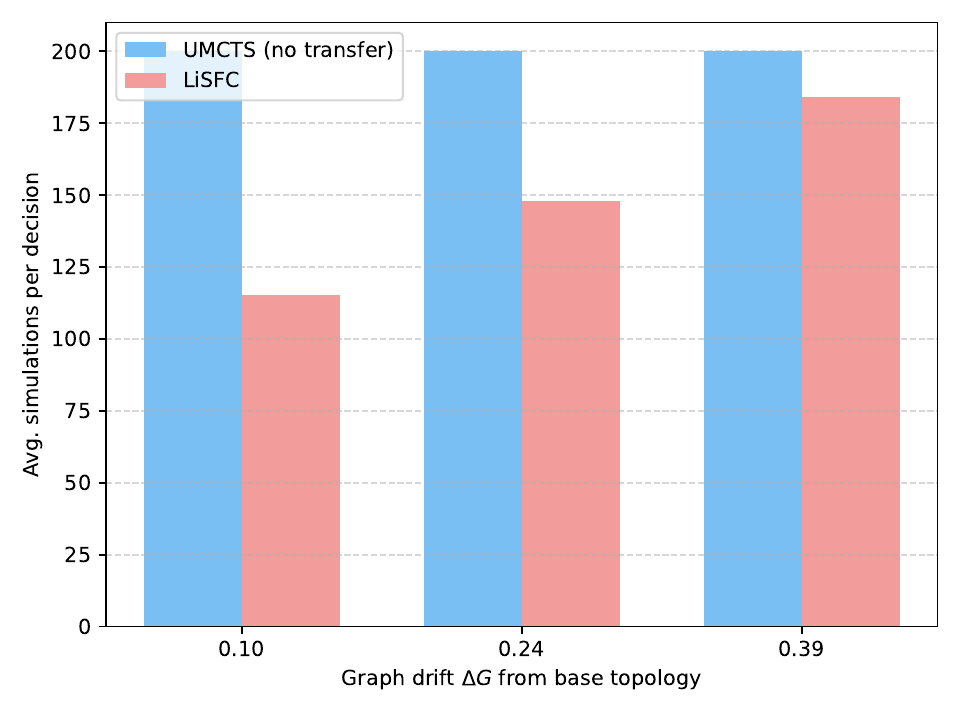}
  \caption{Average MCTS simulations per decision as a function of graph drift \(\Delta G(G_0,G_k)\) between the base topology \(G_0\) and perturbed topologies \(G_1,G_2, G_3\). LiSFC adapts its reuse of past statistics based on estimated MDP distance, reducing search effort for small drift while gracefully reverting to standard UMCTS for large drift.}
  \label{fig:simulations}
  % \vspace{-0.5cm}
\end{figure}

Fig.~\ref{fig:simulations} plots the average number of MCTS simulations per decision for UMCTS and LiSFC as a function of graph drift \(\Delta G(G_0,G_k)\). For small drift (from \(G_0\) to \(G_1\)), LiSFC reuses subtrees and Q-estimates from \(\mathcal{M}_0\), corrected by the transferable aUCT bound using the estimated distance \(\widehat d(\mathcal{M}_0,\mathcal{M}_1)\). This leads to a noticeable reduction in required simulations to reach comparable blocking and P95 delay as in Scenario~1. As the drift increases (from \(G_0\) to \(G_2\)), the estimated distance grows and the aUCT bound becomes looser; LiSFC correspondingly reduces reliance on old statistics and behaves increasingly like UMCTS, which is reflected in the convergence of simulation counts. In all cases, LiSFC does not perform worse than UMCTS in terms of blocking or delay, consistent with the theoretical guarantee that the transfer bias is controlled by \(d(\mathcal{M},\mathcal{M}_i)\).

\subsection{Scenario~3: Cross-task convergence with and without lifelong transfer}

The previous scenarios focused on steady-state performance and average search effort. We now compare the \emph{online convergence} of LiSFC against a non-transfer UMCTS baseline across multiple CPN configurations. Starting from the base topology $G_0$, we first warm up the knowledge base by running LiSFC-Search on $G_0$ with a stream of SFC requests. We then construct three perturbed graphs $G_1,G_2,G_3$ using the upgrade, degraded, and mixed perturbations described in Section~\ref{sec:exp-setup}, and compute their graph drifts $\Delta G(G_0,G_k)$ using our metric from Section~\ref{sec:graph-drift}. For each configuration $G_k$, we process a sequence of SFC requests and record the \emph{running} blocking probability after each decision for both UMCTS and LiSFC.

\begin{figure}[t]
  \centering
  \includegraphics[width=0.95\columnwidth]{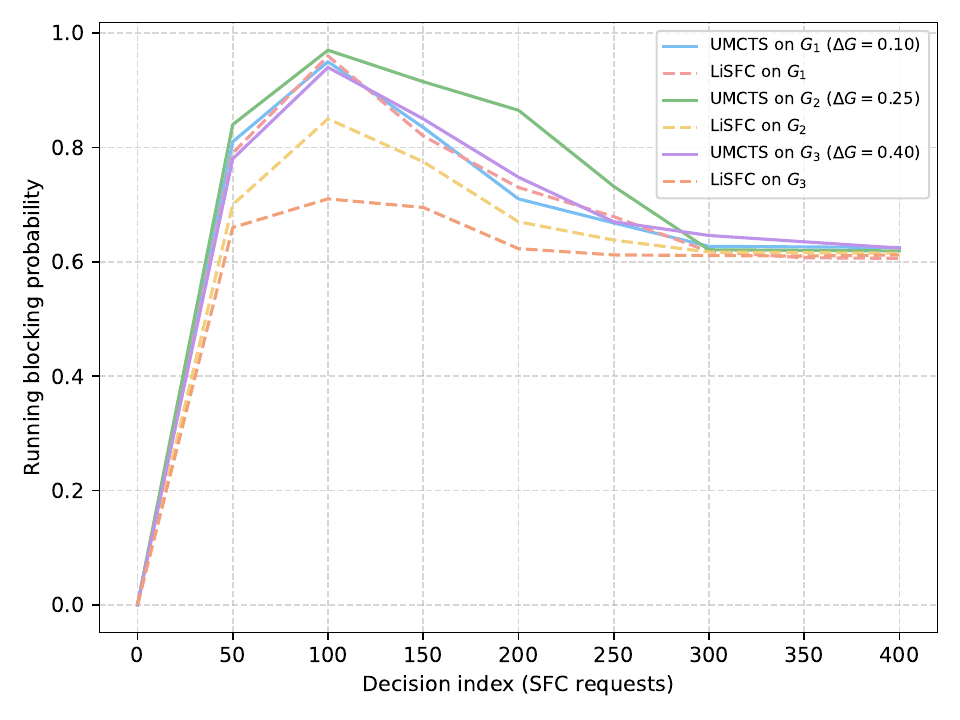}
  \caption{Running blocking probability of LiSFC and vanilla UMCTS as a function of the decision index on three perturbed CPN configurations $G_1,G_2,G_3$. Each configuration has a different graph drift $\Delta G(G_0,G_k)$ from the base topology $G_0$. LiSFC converges faster and to lower blocking on tasks with small drift, while gracefully degrading towards UMCTS behavior as the drift increases.}
  \label{fig:convergence-tasks}
% \vspace{-0.5cm}
\end{figure}

Figure~\ref{fig:convergence-tasks} plots the resulting convergence curves on $G_1,G_2,G_3$. Each task yields two curves (UMCTS and LiSFC). On the upgraded topology $G_1$ with small graph drift, LiSFC starts from a relatively low blocking level, quickly stabilizes, and consistently outperforms UMCTS, reflecting effective reuse of search statistics and subtrees transferred from $G_0$. As the drift increases for $G_2$ and $G_3$, the initial advantage of LiSFC becomes smaller and the convergence behavior gradually approaches that of UMCTS. This trend is consistent with our distance-based design: the transferable aUCT bonuses become looser when $\Delta G(G_0,G_k)$ (and thus the MDP distance) grows, causing LiSFC to rely less on prior knowledge and behave more like a non-transfer planner. Importantly, LiSFC never exhibits worse long-term blocking than UMCTS in our experiments, supporting the claim that lifelong transfer is beneficial when tasks are similar and benign when they are not.

Overall, these two scenarios suggest that LiSFC can exploit the structure of cloud--network convergence in two complementary ways. Across drifting CPN configurations, it uses graph drift and MDP distance estimates to safely reuse planning experience from historical tasks, reducing search effort when drift is small and gracefully falling back to non-transfer MCTS when drift is large. A more extensive evaluation on realistic 5G/6G C-RAN and edge--cloud topologies is an important direction for future work.

\section{Conclusion}
We proposed LiSFC, a Lipschitz lifelong framework for dynamic SFC placing–scheduling in converged cloud–network infrastructures. By modeling SFC planning in computing power networks as a family of related MDPs and introducing a graph-drift metric that upper-bounds LiZero's MDP distance, LiSFC-Search can safely reuse MCTS statistics across evolving cloud–network configurations while remaining robust to topology and load shifts. Preliminary results on synthetic CPN scenarios show that LiSFC reduces SFC blocking and tail E2E delay compared to non-transfer MCTS and learning-only baselines, suggesting that lifelong, distance-aware planning is a promising primitive for AI/ML-based orchestration in cloud–network convergence.

\nocite{*}
\bibliographystyle{IEEEtran}
\bibliography{ref}

\end{document}